\documentclass[aps,twocolumn, superscriptaddress, nofootinbib, longbibliography] {revtex4-1}

\usepackage{comment}
\usepackage[usenames]{color}
\usepackage{xcolor}
\usepackage{upgreek}
\usepackage{amsmath}
\usepackage{amssymb} 
\usepackage{graphicx}
\usepackage[utf8]{inputenc}
\usepackage{physics}
\usepackage{empheq}
\usepackage{braket}
\usepackage{siunitx}
\usepackage{float}
\usepackage{tcolorbox}
\usepackage{enumitem}
\usepackage{placeins}
\usepackage[edges]{forest}

\usepackage{microtype}

\usepackage[T1]{fontenc}
\usepackage{inconsolata} 
\usepackage{xcolor}
\usepackage{listings}
\lstdefinestyle{verilog-default}{
  language=Verilog,
  basicstyle=\fontsize{8}{9.5}\ttfamily\selectfont,
  keywordstyle=\color{blue}\bfseries,
  commentstyle=\color{gray}\itshape,
  stringstyle=\color{orange},
  numbers=left,
  numberstyle=\tiny\color{gray},
  breaklines=true,
  showstringspaces=false,
}

\usepackage{xcolor} 

\DeclareMathSizes{10}{10}{10}{10}

\begin{document}
\title{Synthesis of mass-spring networks from high-level code descriptions} 
\author{Parisa Omidvar}
\affiliation{AMOLF, Science Park 104, 1098 XG Amsterdam, the Netherlands}
\author{Marc Serra-Garcia}
\affiliation{AMOLF, Science Park 104, 1098 XG Amsterdam, the Netherlands}

\date{\today}

\begin{abstract}
Structural nonlinearity can be harnessed to program complex functionalities in robotic devices. However, it remains a challenge to design nonlinear systems that will accomplish a specific, desired task. The responses that we typically describe as intelligent---such a robot navigating a maze---require a large number of degrees of freedom and cannot be captured by traditional optimization objective functions. In this work, we explore a code-based synthesis approach to design mass-spring systems with embodied intelligence. The approach starts from a source code, written in a \emph{mechanical description language}, that details the system boundary, sensor and actuator locations, and desired behavior. A synthesizer software then automatically generates a mass-spring network that performs the described function from the source code description. We exemplify this methodology by designing mass-spring systems realizing a maze-navigating robot and a programmable lock. Remarkably, mechanical description languages can be combined with large-language models, to translate a natural-language description of a task into a functional device.

\end{abstract}

\maketitle

\section*{}
\section{Introduction}
Nonlinear elastic systems, from origami linkages \cite{meng2021bistability, yasuda2017origami} to buckling beams~\cite{kwakernaak2023counting},  have demonstrated remarkable information-processing capabilities, including the realization of logic operations~\cite{mei2023memory, el2022mechanical, ion2017digital, drotman2021electronics, song2019additively}, programmable matrix-vector multiplication, speech recognition and the nonvolatile information storage~\cite{chen2021reprogrammable}.
These results have radically shifted the perspective on mechanical nonlinearity: from something to avoid entirely~\cite{postma2005dynamic}, to an essential resource for the realization of novel functionalities~\cite{jiao2023mechanical, barri2021multifunctional, shan2015multistable, mohammadi2021flexible, louvet2025reprogrammable}. An important area of application for mechanical nonlinearity is to embody intelligent responses in soft robotic devices. Recent examples have fabricated structures capable of autonomous navigation and object sorting~\cite{kamp2025reprogrammable}. While these results are promising, designing an elastic structure that exhibits intelligent behavior is still a research project in itself; we lack tools that can systematically take an engineering specification and translate it into a functional design~\cite{meeussen2025new, fard2025embodying}. Traditional computational optimization, for example, requires a well-defined misfit function, such as a transmitted displacement or energy~\cite{louvet2025reprogrammable, chen2024intelligent, zeng2023inverse}. This approach is limited by the expressivity of the target misfit function, as intelligent systems respond differently to different inputs, with the space of possible responses increasing exponentially with the dimension of the input space.

\begin{figure}[t]
    \centering
    \includegraphics[width=\columnwidth]{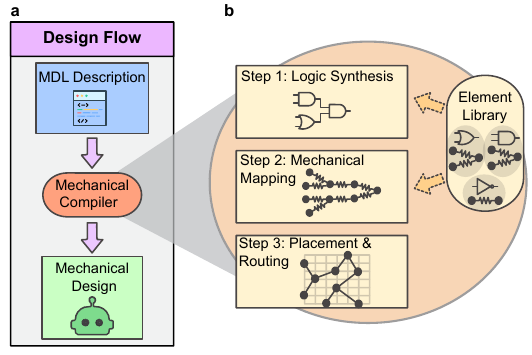}
    \caption{\textbf{Automated mechanical synthesis design.} \textbf{a}, Design flow for mechanical synthesis. The designer specifies the desired behavior of the system using a Mechanical Description Language (MDL). This description is interpreted by a mechanical compiler which generates the design accordingly. \textbf{b}, The mechanical compiler first converts the MDL code into a gate-level netlist, then maps the list to its physical equivalent using elements from the library made of bistable-masses and variable-stiffness couplings, and finally assigns the spatial coordinates while satisfying geometric constraints and prescribed input/output locations. } 
    \label{fig:DesignFlow}
\end{figure}

In this paper, we explore a code-based approach to synthesize mechanical systems with embodied intelligence (Fig. \ref{fig:DesignFlow}a). This approach is inspired by the digital synthesis methods ~\cite{gajski1996principles, manodigital, wolf2013yosys} used in electrical engineering. In digital synthesis, complex functionalities such as a counter or a microprocessor are generated from a higher-level source code description, written in a Hardware Description Language (HDL). Then, the synthesizer translates the code into a functional design.  HDLs facilitate the testing and reuse of designs, as code can be simulated easily before fabrication, and the same source can be compiled into FPGAs or application-specific integrated circuits. We will explore the use of HDLs for mechanics---which we refer to as \emph{Mechanical Description Languages} (MDLs), by synthesizing nonlinear mass-spring networks from a code description. Nonlinear mass-spring networks can be used as a model for generic elastic structures, abstracting away geometric details yet retaining sufficiently expressivity to represent functionalities ranging from speech recognition~\cite{bohte2025general} to digital logic~\cite{serra2019turing}. To design the mass-spring networks, our mechanical compiler interprets these behavioral descriptions in three stages (Fig. \ref{fig:DesignFlow}b): First, the behavior is mapped to a digital circuit, consisting of logic gates and memory elements; then, the digital circuit is translated into a mass-spring network; finally, the mass-spring network is placed inside the device structure and routed (interconnected)---satisfying the specified embodiment. Because HDLs are text-based descriptions of a desired behavior, they are easy to interface with Large Language Models. We end our work by showing how a recent model (GPT-5) is capable of generating functional MDL code from a high-level, plain-language description of the function.

Our paper is structured as follows: We will start by a description of our MDL synthesis workflow, including an overview of the language structure, the mass-spring building blocks into which the design is compiled, and the compilation process. Then, we will synthesize and numerically simulate two example designs generated via MDL: A robot that navigates a maze and a numerical passcode lock, generated via MDL synthesis. Then, we will show how both examples can be generated with Large Language Models from a plain text descriptions of the desired function. Finally, we will discuss our results, including limitations and further work.



\section{The Mechanical Description Language and synthesis design flow}

\subsection{Code structure}
\begin{figure}[t]
    \centering
    \includegraphics[width=\columnwidth]{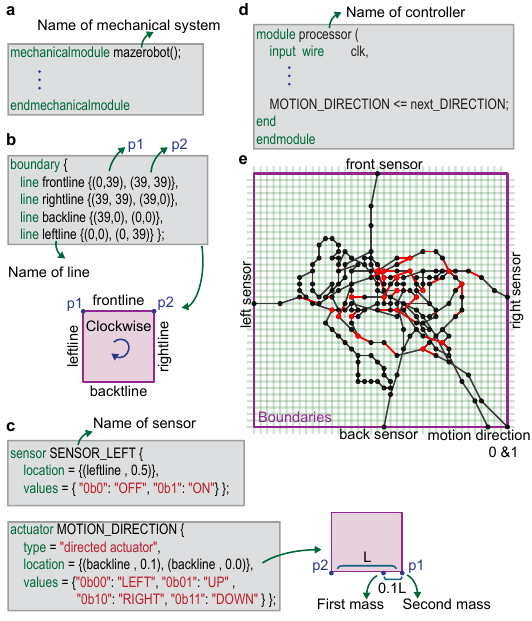}
    \caption{\textbf{Mechanical description language and generated mass-spring network} \textbf{a} The keyword \emph{mechanicalmodule} is used to identify and name a component---here named  \textit{mazerobot} in correspondence with example I. 
    \textbf{b} The shape (boundary) of the element is defined as a rectangle, composed of a set of straight lines---identified by the keyword \emph{line}. The lines are  specified by a name and two integer endpoints, and listed in clockwise order---enabling the algorithm to identify the inside/outside of the robot. 
    \textbf{c} The location, and type of the sensors and actuators is specified through \emph{sensor} and \emph{actuator} elements. The location is specified as a coordinate along a specific wall, while the value field maps sensor measurements and actuator actions to corresponding binary values. For the purpose of this example, we include only one type of sensor and one type of actuator, but in principle large libraries could be constructed.
    \textbf{d} The next block of the code defines the logical behavior of the system, using the Verilog syntax.
    \textbf{e} An example synthesized mass-spring network generated from the MDL code using the mechanical compiler. The circles represent the masses in the network (with the color corresponding to the biasing force). The lines represent interacting potentials as described in Fig.~\ref{fig:LibraryElements}, with the red bars indicating negative interaction potentials. } 
    \label{fig:PLace&Route}
\end{figure}

The MDL code in this work (Fig.~\ref{fig:PLace&Route}a-d) is structured in three sections. The first section describes the geometric boundary of the device. The second section describes the locations of the sensors and actuators by which the device interacts with the environment, and the third section describes the behavior of the device, i.e., how the internal state is updated in response to the input from the sensors, and how the actuators depend on the internal state. The update algorithm is based on the existing Verilog HDL, while the device boundary and actuator location are described with a custom syntax. This allows us to rely on existing Verilog synthesis tools to generate the control part of the design.
The proposed language constrains the embodiment (shape) and the behavior of the device, it does not constrain the implementation; Here, we will compile our design into a network of masses and springs (Fig.~\ref{fig:PLace&Route}e), the same code could in principle be compiled into a fluidic or electric circuit, or to a nanomagnet-based robotic device~\cite{cui2019nanomagnetic}.

\begin{figure*}[t]

    \includegraphics[width=\textwidth]{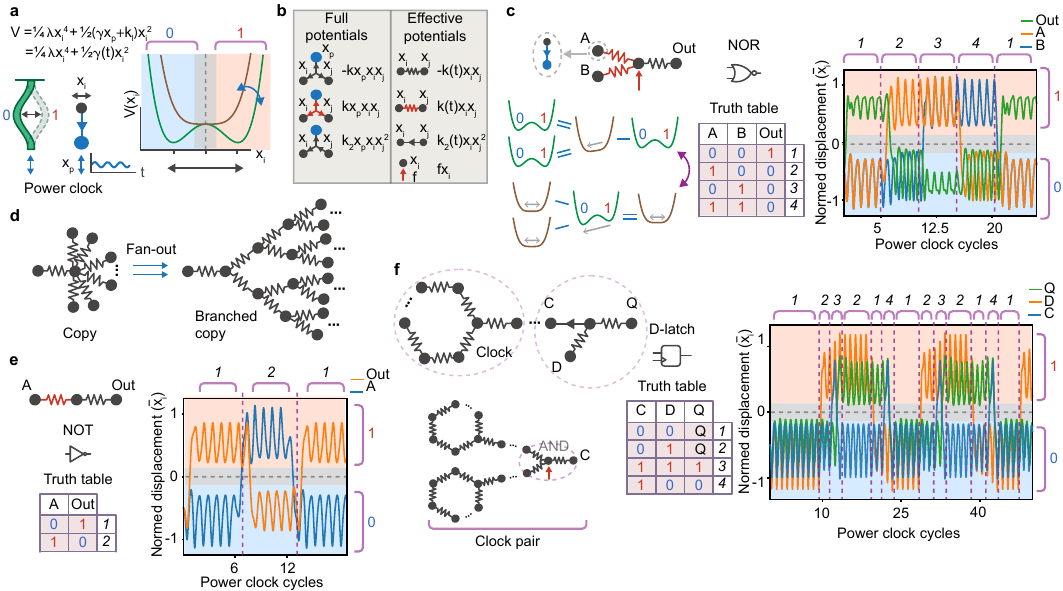}
    \caption{\textbf{Mechanical logic building blocks.} 
    \textbf{a}, The basic building block is a mass, subject to a potential that can be switched between single and double well configurations. The \textit{power clock} (blue mass) modulates the potential from bistable to monostable, enabling state switching.
    \textbf{b} Pairs of masses are connected through nonlinear potentials, whose energy form depends on the displacement of each mass ($x_i$, $x_j$) and the \textit{power clock} ($x_p$). Because the \textit{power clock} is externally prescribed, the coupling potentials effectively behave as a time-dependent, pairwise interaction. Depending on the specific energy form, this effective interaction can be linear or nonlinear.
    \textbf{c} Top left: The universal NOR gate design, implemented by in a four-mass system using negative effective springs and a local biasing potential. Bottom left: Numerical simulation of the time-evolution of the on-site potentials acting on the masses. When the coupling of the inputs and the intermediate mass are strong, the inputs are subject to a double well potential while the intermediate mass is monostable---allowing it to follow the inputs. As the information propagates one site forward, the coupling weakens between the input and intermediate masses while strengthening between the intermediate and output masses. This cycle is then repeated. Right: The displacement amplitude of the inputs and outputs over the \textit{power clock} cycles, exhibiting a NOR gate response. The positive (negative) displacements shows that the corresponding mass takes $1$ ($0$) logical state.
    \textbf{d} When multiple inputs are connected to a single output, we use a branching mechanism constructed from buffer gates. 
    \textbf{e} A NOT gate is similar to a buffer, but it inverts the input signal through a negative effective coupling. The graph shows a numerical simulation of the input and output displacements as the \textit{power clock} cycles.
    \textbf{f}, Clock circuit driving a D-latch. We use a loop of masses of length $N$ to create a periodic clock signal. The D-latch consists of four masses: Two inputs (clock $C$ and data $D$), an intermediate mass, and an output. The clock input $C$ is connected to the intermediate mass via a nonlinear interaction, causing the potential to switch from monostable to bi-stable when the clock ticks. The data input $D$ biases the intermediate mass towards $0$ or $1$, causing a specific value to be latched when the single well-double well transition is triggered by a clock tick. Longer effective clocks can be constructed by connecting multiple, co-prime loops through an AND gate. The plot presents a numerical simulation of a clock-latch tandem. The latch input $D$ (orange) is copied into the output $Q$ (green) when the clock $C$ input ticks (blue).
    } 
    \label{fig:LibraryElements}
\end{figure*}

\subsection{Building blocks}

The basic building block of the mass-spring system will be a mass subject to a nonlinear potential, that can be switched between a double-well and single-well configuration (Fig.~\ref{fig:LibraryElements}a). In the double-well configuration, the mass presents two stable equilibria, corresponding to the logical states of $0$ (low) and $1$ (high)---storing a single bit of information. Switching to a single-well configuration allows the information on the mass to be erased and rewritten. Although the scope of this work is idealized mass-spring systems, in elastic systems double well potentials can be realized with a beam under compression. In a high-compression setting, the beam will buckle into either of two stable configurations---realizing the double-well potential, while in the low-compression scenario the beam will present a single stable configuration. 
Switching between double-well and single-well setting is governed by an additional degree of freedom, that we refer to as the \textit{power clock}---and intuitively captures the 'boundary compression' of the buckling beam. This element provides both the timing (clocking) of the system, as well as the energy that drives state transitions. Throughout this work, we will prescribe the \textit{power clock} to follow a harmonic displacement (see Fig.~\ref{fig:LibraryElements}a). The nonlinear interaction between the mass and the \textit{power clock} is governed by a quadratic coupling potential~\cite{bohte2025general} (Fig.~\ref{fig:LibraryElements}b). In an experimental setting, this \textit{power clock} could be provided, for example, by a pneumatic oscillator~\cite{van2022fluidic} or an external magnetic field~\cite{cui2019nanomagnetic}.

Information processing requires signals to propagate between masses. We achieve this by connecting the masses through nonlinear interactions that are also modulated by the \textit{power clock} (Fig.~\ref{fig:LibraryElements}b). Since the displacement of the \textit{power clock} varies with time, the corresponding effective interaction between connected masses becomes time-dependent (Fig.~\ref{fig:LibraryElements}b). In this work, we will consider three types of emergent connections between the masses: a positive (negative) linear spring, and an asymmetric nonlinear coupling element. In a practical scenario, variable-stiffness coupling can be realized via variable-stiffness compliant mechanisms \cite{kuppens2021monolithic}. From the aforementioned bi-stable elements and coupling potentials, we can construct the digital computing building blocks that will be combined to realize the control algorithm of the system, as defined in the MDL source code. Although in principle we can realize a diversity of digital elements using mechanical components, here we will operate with a minimal set of five building blocks: a universal NOR gate, a NOT gate, a buffer, a D-latch and clock generating element. Because the NOR gate is universal, we will be able to realize any digital function by combining these elements.

To construct the \emph{NOR gate}, we connect two input masses (labeled \textit{A} and \textit{B}) to an internal mass through a negative coupling. The internal mass is then connected to the output mass (labeled \textit{Out})---which, in a circuit, will also correspond to an input mass of the subsequent logic gate (see Fig.~\ref{fig:LibraryElements}c). The internal mass is coupled to the \textit{power clock} in an opposite way to the input/output masses: When the input and output are transitioning from the bistable to the monostable state, the internal mass transitions from the monostable to the bistable state. This type of out-of-phase interaction has been experimentally realized by connecting opposite ends of a beam to a moving frame \cite{omidvar2025racetrack}. The direction of signal propagation is determined by the two distinct phases of the coupling elements, which alternate between high-stiffness and low-stiffness states through the \textit{power clock} cycle. Because one of the internal coupling springs is negative, the output of the NOR gate responds inversely to the inputs: When both inputs are $0$, the output becomes $1$, and vice versa. In the case where one of the inputs is $0$ and the other is $1$, the input forces compensate. To induce a NOR response,  we add an extra force (see Fig.~\ref{fig:LibraryElements}b) on the intermediate mass, that disambiguates this result by biasing the output towards $0$.

In complex digital circuits, a gate output may need to serve as the input for multiple gates. When connecting a single output to a large number of inputs, we observed that sometimes circuits did not operate as expected. This is due to the excessive stiffness load from the multiple inputs on the single output. To address this issue, we introduce a \emph{buffer gate}, that copies an input into the output. With a buffer gate, we can construct tree-like structures to fan-out digital signals (Fig.~\ref{fig:LibraryElements}d). When one of the springs in the buffer is replaced by a negative spring, the buffer becomes a \emph{NOT gate} (Fig.~\ref{fig:LibraryElements}e), which is used to invert the input signal. 

The NOR, buffer and NOT gates are sufficient for realizing combinatorial logic. However,
for sequential logic, where the output depends not only on the inputs but also on the past states of the system, two additional building blocks are necessary: First, a clock that synchronizes the state updates; second, a memory element---called a \emph{D-latch}---that stores a state, which is updated in response to clock ticks. 

We construct the clock by forming a closed loop of $N$ connected masses (Fig.~\ref{fig:LibraryElements}f). We set an initial displacement of $1$ for one of the masses, while the rest of the masses in the loop are initialized to $0$. When the \textit{power clock} is driven, the bit set to $1$ continuously circulates through the loop, reaching the output site once every $N$ \textit{power clock} cycles. When designing a circuit, we set the length $N$ of the loop to ensure that the information has time to propagate through the slowest combinatorial path before being latched. In general, this dictates the use of large $N$ values. To avoid using very long clock circuits, we combine two clocks with co-prime numbers of masses, using an AND gate. In this case, the effective length of the clock pair is the product of both clocks' length. 

The \emph{D-latch} (see Fig.~\ref{fig:LibraryElements}f) contains two input masses, an intermediate mass and an output mass. The input masses are the clock and data inputs respectively (labeled \textit{C} and \textit{D}). The data input is linearly connected to the intermediate mass, while the clock input is coupled to the intermediate mass through a nonlinear interaction similar to the one connecting each mass to the \textit{power clock}. Because of this nonlinear interaction, the intermediate mass exhibits bistability only when the clock signal is positive---causing data to be latched only when the clock ticks. Since the state transition occurs at a specific moment in time, the D-latch behaves as a mechanical equivalent of a digital flip-flop. 

\subsection{Mechanical Compiler}
The goal of the mechanical compiler is to realize devices that embody the behavior described in the MDL code, by combining the building blocks described in the previous section. The compiler starts by identifying the three sections of the code. 
The behavioral block then converted to a network of logic gates using an open-source digital synthesis tool \cite{wolf2013yosys}. The output of the digital synthesis tool is then modified to prevent more than two inputs to be connected to the same output, by inserting buffer gates as needed. Then, the compiler replaces each logic gate (NOR, NOT, Buffer and D-latch) by the corresponding mechanical building block. If the circuit is sequential, the compiler automatically generates the clock circuit based on a synthesis parameter that defines its length---in principle, one could determine the minimum code length from the synthesis report, but here we are defining it manually. Once the network of masses and springs has been generated, the compiler identifies the device boundary geometry, and the positions of the actuator and sensor masses---which are treated as design constraints. Then, it uses a place-and-route algorithm to distribute the unconstrained masses (those that are not sensors and actuators) inside the device, to minimize wiring distances and crossings (see Fig.~\ref{fig:DesignFlow}b and the Appendix for details).

\begin{figure*}[t!]
    \includegraphics[width=\textwidth]{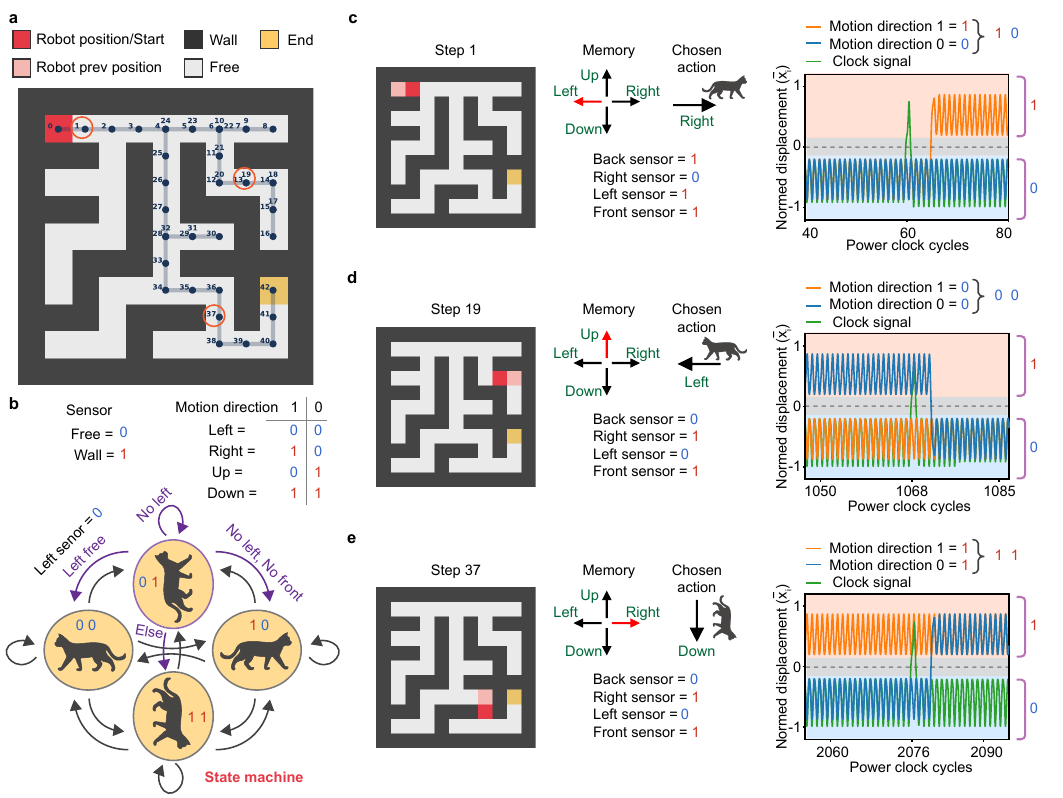}
    \caption{\textbf{Numerical simulation of a maze-following robot.} \textbf{a} Path taken by a robot generated from a MDL description. The mass-spring model (Fig.~\ref{fig:PLace&Route}e) is simulated via time-integration, subject to external sensor forces that depend on the position within the maze.
    \textbf{b} The robot response is governed by a finite-state machine (FSM), with four states corresponding to the four possible directions of motion \textit{left, right, up} and \textit{down}. These states are mapped to actuator outputs of $00$, $10$, $01$ and $11$ respectively. The FSM updates its direction based on the sensor inputs, that detect the walls around the robot.
    \textbf{c}, The initial state is \textit{left} ($00$). Sensor inputs indicate that all the directions are blocked except \textit{right}. Therefore, The system switches to \textit{right} motion, updating its outputs to $10$ after receiving the clock signal.
    \textbf{d}, In step 19, the FSM has a directional state of \textit{up} ($01$) and the sensor inputs indicate that \textit{right} and \textit{front} are blocked. Following the FSM update rule, the machine turns \textit{left}, ending in state $00$.   
    \textbf{e}, In step 37, the directional state of the machine is \textit{right} ($10$). Based on the sensor inputs the FSM updates its state to \textit{down} motion by switching to the $11$ state.
    } 
    \label{fig:maze}
\end{figure*}

\section{Examples}

\subsection{Example I: Maze-following robot}

A paradigmatic example of embodied intelligence is a robot that can find the exit of a maze. Here, we explore the automatic generation of such a robot. The body of the robot is a rectangle (Fig.~\ref{fig:PLace&Route}b); at the midpoint of every boundary, we place a sensor that detects the presence of a wall (Fig.~\ref{fig:PLace&Route}c). Although here the sensors are modeled as a wall-sensitive force, in an actual device these could be realized through a bistable mechanical element. In a corner of the robot, we program an actuator whose motion will depend on a logical variable (Fig.~\ref{fig:PLace&Route}c). Finally, we write a behavioral description indicating how the direction of motion should be updated in response to the sensor input  (Fig.~\ref{fig:PLace&Route}e). The full MDL code is provided in the appendix.

The mechanical compiler uses this description code and the library elements described in Fig.~\ref{fig:LibraryElements} to generate a network of 187 interconnected masses. The compiler places these masses within a square-shaped boundary while preserving the specified locations of the inputs and outputs (Fig.~\ref{fig:PLace&Route}e).


We numerically simulate the network within a maze environment, as illustrated in Fig.~\ref{fig:maze}a. The interactions between the robot and the maze walls are modeled as applied forces on the four input masses (sensors). When the robot advances to a new position, these input forces are updated to reflect the new contact configuration with the maze boundaries.
The robot operates as a finite-state machine (Fig.~\ref{fig:maze}b)---generated automatically from the code description---that determines the next directional motion state based on its current directional state and local sensory inputs. The navigation algorithm follows the left-wall-following rule. At each time step, the robot evaluates the feasibility of moving in each possible direction, following a fixed priority order that depends on its current directional state.
The movement rules can be summarized as follows.

\begin{itemize}[nosep] 
\item If moving \textit{left}: the priority of motion is \textit{down} $\rightarrow$ \textit{left} $\rightarrow$ \textit{forward} $\rightarrow$ \textit{right}
\item If moving \textit{up}: the priority of motion is \textit{left} $\rightarrow$ \textit{up} $\rightarrow$ \textit{right} $\rightarrow$ \textit{down}
\item If moving \textit{right}: the priority of motion is \textit{up} $\rightarrow$ \textit{right} $\rightarrow$ \textit{down} $\rightarrow$ \textit{left}
\item If moving \textit{down}: the priority of motion is \textit{right} $\rightarrow$ \textit{down} $\rightarrow$ \textit{left} $\rightarrow$ \textit{up}
\end{itemize}

The movement output is encoded as a two-bit signal, where the first bit denotes the motion axis ($0$ for horizontal, $1$ for vertical), and the second bit indicates direction along that axis ($0$ for \textit{up/left}, $1$ for \textit{down/right}). In the numerical simulation, we read out this actuated direction and update the coordinates of the robot. Figures~\ref{fig:maze}c,d and e illustrate the sensor input and evolution of actuator state through different states of the maze-solving. 

\begin{figure}[t]

    \includegraphics[width=\columnwidth]{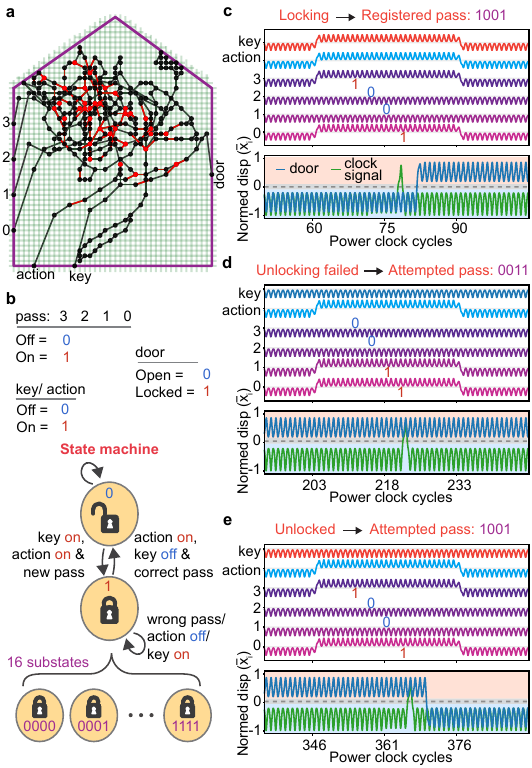}
    \caption{\textbf{Synthesis of one-time code lock.} \textbf{a}, The synthesized mass-spring network embodied within a defined boundary. The system has inputs representing the four-digit binary passcode, with masses representing \textit{key} and \textit{action}, as well as a \textit{door} output mass.
    \textbf{b}, The system operates with two primary states of \textit{unlocked} ($0$) and \textit{locked} ($1$). When \textit{locked} by applying the passcode while both \textit{key} and \textit{action} are set to state $1$, the system falls into one of sixteen sub-states of \textit{locked}, corresponding to all possible passcode combinations. To open the lock, the registered passcode together with the \textit{action} should be pressed.
    \textbf{c}, Locking process. Top: We apply a force on the  \textit{action} and  \textit{key} masses to set the state $1$ while applying a desired passcode ($1001$). Bottom: Initially the system is at  \textit{unlock} state, with  \textit{door} mass being at the negative stable state ($0$), after the clock signal, the passcode is registered and the \textit{door} moves to positive stable state ($1$). Then, the system is \textit{locked}.
    \textbf{d}, Unlocking attempt with an incorrect passcode. We set the \textit{action} key to positive stable state ($1$), along with an incorrect passcode ($0011$). The output stays in the positive stable state, and the system remains \textit{locked}.
    \textbf{e}, Successful unlocking. Using the correct passcode ($1001$) together with forcing \textit{action} input to ($0$), causes the \textit{door} to change state from $1$ to $0$, and the system becomes \textit{unlocked}. 
    } 
    \label{fig:lock}
\end{figure}

\subsection{Example II: Numerical combination lock}

As a second example, we synthesize a numerical combination lock, inspired by reprogrammable gym locks. When the lock is closed, it remembers the numerical combination that was pressed during closing. This numerical combination is required to re-open the lock. Figure~\ref{fig:lock}a shows the synthesized network of 311 interconnected masses corresponding to this example (see the Appendix for the corresponding MDL code). 
The lock contains six input degrees of freedom, representing the \textit{key} and \textit{action} input buttons (masses), along with four inputs for entering the passcode. Each input takes a binary value of $0$ or $1$. 

The lock functions as a finite-state machine with an \textit{unlocked} state, and 16 \textit{locked} states---each of them corresponding to a specific four-bits passcode (Fig.~\ref{fig:lock}b). Figure~\ref{fig:lock}c illustrates the case where both the action and key inputs are set to $1$, and the passcode encodes the combination $1001$.
Since both the action and key buttons are pressed simultaneously, the passcode is registered as the new access code, and the state of the system transitions from \textit{unlocked} to \textit{locked} after the next clock signal.
In Fig.~\ref{fig:lock}d, an unlocking attempt with an incorrect passcode ($0011$) fails to open the lock, and the system state remains \textit{locked}.
In contrast, as shown in Fig.~\ref{fig:lock}e, the subsequent attempt with the correct passcode ($1001$) results in a successful unlocking, with the \textit{action} and \textit{key} inputs set to $1$ and $0$, respectively.

\section{Design Automation with Large Language Models }
MDL uses text to represent mechanical functionalities. As such, it is straightforward to interface with LLMs. To evaluate whether an LLM can generate an MDL code from a high-level, natural-language specification, we use the OpenAI GPT-5 API (version gpt-5-2025-08-07) to generate the prior examples. In the API call, we provide a system prompt that defines the syntax and semantics of our MDL, indicating how geometry, sensing and actuation are encoded (see the Appendix for the full system prompt). This system prompt is the same for every generated device, and is not meant to be altered by the end-user. In the user prompt, we describe the behavior that we intend to generate.

User prompt for Example I:

\lstdefinestyle{diff}{ basicstyle=\ttfamily\footnotesize, breaklines=true, 
frame=single, backgroundcolor=\color{gray!5},
breakindent=0pt, 
columns=flexible,
keepspaces=true, 
moredelim=**[is][\color{red}]{@@}{@@}
}

\UseRawInputEncoding
\begin{lstlisting}[style=diff]
Write MDL code for a simple maze-solving robot.
Make a 39*39 square boundary with four sides.
Add four sensors (one per side, ON when there is a wall) and one actuator that controls movement (LEFT, UP, RIGHT, DOWN).
Make the robot decide its next direction assuming the maze does not have any loops.
\end{lstlisting}

User prompt for Example II:

\UseRawInputEncoding
\begin{lstlisting}[style=diff]
Write an MDL code for a lock mechanism with a 4-bit input, an action button, and a lock/unlock signal. 
When locking (only when the door is unlocked), the applied 4-bit keypad input is stored as the password. When unlocking, the system compares the stored password with the input. If they match, the door opens (0); otherwise, it stays locked (1).
The boundary is a house-shaped pentagon with maximum values of 39*49 and walls of 35 units.
The keypad inputs are located in the left boundary, all other inputs on the bottom boundary and the output on the right boundary.
\end{lstlisting}

Remarkably, the LLM successfully generates a structurally and behaviorally valid MDL code. The mass-spring networks synthesized using these codes behave similarly to those presented in Figs.~\ref{fig:maze} and ~\ref{fig:lock}---although the maze-solver followed the left-hand instead of right-hand rule (see the Appendix for the automatically generated codes and simulations). 
 
\section{Discussion and conclusion}
In this paper, we have explored the automated generation of mass-spring networks with embodied intelligence from a source-code specification through the use of a \emph{mechanical compiler}. The required source code can be produced from a natural-language description using an LLM. This ability can have a significant impact~\cite{whitney1996mechanical, antonsson1997potential}. Currently, designing an elastic structure that exhibits a particular intelligent functionality is a challenging research project. Mechanical synthesis can empower end-users to generate personalized intelligent devices on demand. 

The MDL introduced in this work is extensible, enabling the incorporation of novel sensors, actuators. Because the language describes the desired embodiment and behavior of the system, but not the implementation, design decisions such as whether the control logic should be performed in the mechanical or electronic domain can be automatically determined by the compiler based on objectives of energy efficiency, cost and size. When newer compilers are produced (for example, including improved logic elements), designs can be generated automatically by simply recompiling existing code. 

This work has focused on the \emph{toy example} of generating mass-spring networks according to digital update rules. Future works should move beyond these simple demonstrations and address the generation of three-dimensional geometries, as well as the incorporation of precise mechanical parameters, such as sensor forces and actuator velocities and accelerations.

\section*{Acknowledgements}
The authors would like to thank Martin van Hecke and Finn Bohte for valuable discussions.

Correspondence can be addressed to Marc Serra-Garcia (m.serragarcia@amolf.nl). 

\bibliography{bib}

\newpage

\appendix

\section*{Appendix}

\section{Methods}

We solved Ordinary Differential Equations
(ODEs) of the mass-spring networks in this paper using a 4th-order Runge-Kutta solver with 400 points per period. 
The total potential energy of the system is expressed as:

\begin{equation}
\begin{split}
V = &\sum_{i}   \frac{\lambda}{4} \, x_i^{4} + \frac{k_l}{2} \, 
x_i^{2}  \\ &+ \frac{\gamma}{2} \,x_i^{2}(a_{0,i}+ a_{1,i} x_p) \\ &
 + q_i x_i \\ &
\sum_{i,j} - \frac{nc_{i,j}(d_{0,i,j} +d_{1,i,j} x_p)}{2}\ x_ix_j^{n}  
\end{split}
\label{eq:potential}
\end{equation}

The equation of motion for each mass is given by:

\begin{equation}
F_i(t) = m \ddot{x_i} + b \dot{x_i} + \sum_{} \frac{\partial V}{\partial x_i}
\end{equation}

where $x_i$ denotes the displacement of the $i^{\text{th}}$ mass and $F_i(t)$ is the sum of the forces acting on the $i^{\text{th}}$ mass. $\lambda$, $k_l$, $\gamma$, $c$, $m$ and $b$ represent the duffing nonlinearity, the quadratic constant, the quadratic coupling, the linear coupling, mass and damping coefficients, respectively. In our simulations, we consider the values as $\lambda =1$, $k_l =1.5$, $\gamma =-2$ and $c_{i,j}= c_{j,i} =0.5$, $m = 0.05$ and $b =0.25$. These values ensure that the equations are overdamped. Whenever any mass is biased we use $q_i = 1$ to add a negative force on the corresponding mass. In addition, whenever we have a negative coupling, we set the coupling value as $c_{i,j} = c_{j,i} =-0.5$. If the effective interaction is nonlinear $n = 2$, $c_{i,j} = -0.5$, and $c_{j,i} = 0$, otherwise $n = 1$. This interaction is asymmetric.

In Eq. \ref{eq:potential}, the \textit{Power clock} mass ($x_p$) follows a sinusoidal trajectory given by $x_p = 1 + sin(\omega t)$. We consider two phases for both the masses and the variable couplings with respect to their interaction with $x_p$, characterized by the coefficients $a_0$, $a_1$, $d_0$, and $d_1$. For in-phase cases, the coefficients are set as $a_0 = d_0 = 0$ and $a_1 = d_1 = 1$, whereas for out-of-phase cases we use $a_0 = d_0 = 2$ and $a_1 = d_1 = -1$.

\section{Synthesis process}

The hardware description block of the MDL is written in Verilog. To compile the Verilog part, we use Yosys, an open-source digital synthesizer (version 0.17+67, git SHA1 01cb02c81, compiled with clang 10.0.0-4ubuntu1 using -fPIC -Os). The synthesized netlist of logic gates generated by Yosys is stored in JSON format.

We implement the place-and-route procedure for the layout of the mass-spring network in Python. First, we use the NetworkX package (version 3.2.1) to construct a graph whose nodes and edges represent the masses and coupling elements, respectively. 
The geometric boundary of the device is then discretized into integer grid points, and the positions of sensors and actuators are mapped to fixed points on the boundary. 
We employ a force-directed algorithm (Fruchterman-Reingold method) to generate an initial continuous 2D layout of the network. This algorithm models the network as a physical system, where edges pull connected nodes together like springs, and nodes repel each other like charged particles. The algorithm iteratively adjusts the node positions until the forces are balanced, producing a well distributed layout. 
Next, the continuous coordinates are projected onto the integer grid by solving a constrained assignment problem using CP-SAT solver from Google OR-Tools (version 9.14.6206), taking into account the fixed points. This process is then repeated, with the fixed points also constrained during the second force-directed layout stage. Finally, we apply a greedy local post-processing step to reduce edge crossings. This algorithm starts from the generated layout and repeatedly scans nearby pairs of non-pinned nodes, evaluates the number of edge crossings before and after swapping their positions, and accepts only swaps that reduce the number of crossings. In each evaluation, only edges incident to the swapped nodes are considered when counting crossings.

During the automated design process, the user specifies several synthesis parameters. The primary parameters include the MDL code and the name of the library (if changed). The clock length, which has a default value of 60, can also be adjusted. For the place-and-route stage, users may specify the number of random seeds and the number of iterations for the two layout stages. If the edge-crossing reduction stage is enabled, the maximum time limit, the number of iterations, and the neighborhood radius used for node swapping can also be defined. The compiler provides default values for these parameters. However, the user can modify them to, for example, extend or shorten the iteration time, increase the maximum runtime, or explore alternative layout configurations. In addition, several optional parameters can be set, such as disabling automatic clock generation, disabling tree-like branching, and disabling the assignment of separate gates to each input/output. If the standard cells in the library are modified, the name of the buffer gate must also be specified when either of the last two options is enabled.

These procedures are executed within a main JupyterLab notebook, which serves as the user interface for the compiler and provides an automated design environment. The library is generated separately on a JupyterLab notebook. Running this notebook creates the library files required for synthesis, making the library easily adjustable. In addition, we implemented the supporting classes and functions within the \textit{Losyspring} custom Python package (short for \textit{Logical Synthesis of Springs}), which provides the underlying design automation tasks.

\section{MDL code for Example I}

\begin{lstlisting}

mechanicalmodule mazerobot();
  //Geometry definition-clockwise on the grid (integer numbers)
   boundary {
      line frontline {(0,39), (39, 39)},
      line rightline {(39, 39), (39,0)},
      line backline {(39,0), (0,0)},
      line leftline {(0,0), (0, 39)} };

  //Sensor definitions (name + location + value) 
  sensor SENSOR_LEFT {
      location = {(leftline , 0.5)}, //0 to (distance counted from point 1)
      values = { "0b0": "OFF", "0b1": "ON"} };
  sensor SENSOR_FRONT {
      location = {(frontline , 0.5)},
      values = { "0b0": "OFF", "0b1": "ON"} };
  sensor SENSOR_RIGHT {
      location = {(rightline , 0.5)},
      values = { "0b0": "OFF", "0b1": "ON"} };
  sensor SENSOR_BACK {
      location = {(backline , 0.5)},
      values = { "0b0": "OFF", "0b1": "ON"} };

  // Actuator (name + location + value) 
  actuator MOTION_DIRECTION {
      type = "directed actuator",
      location = {(backline , 0.1), (backline , 0.0)},
      values = { "0b00": "LEFT", "0b01": "UP" , "0b10": "RIGHT", "0b11": "DOWN" } };

    module mazeProcessor (
        input  wire        clk,
        input  wire        SENSOR_LEFT,  //1 = wall present, 0 = wall absent
        input  wire        SENSOR_FRONT,   
        input  wire        SENSOR_RIGHT,
        input  wire        SENSOR_BACK,
        output reg  [1:0]  MOTION_DIRECTION ); //00=LEFT, 01=UP, 10=RIGHT, 11=DOWN 

    localparam LEFT  = 2'd0;
    localparam UP    = 2'd1;
    localparam RIGHT = 2'd2;
    localparam DOWN  = 2'd3;        
    localparam WALL_PRESENT = 1'b1;
    localparam WALL_ABSENT  = 1'b0;   
    reg [1:0] next_DIRECTION; // next‐state register

    always @(posedge clk) begin        
        case (MOTION_DIRECTION)
        LEFT: begin
            if (SENSOR_BACK  == WALL_ABSENT) next_DIRECTION = DOWN;
            else if (SENSOR_LEFT == WALL_ABSENT) next_DIRECTION = LEFT;
            else if (SENSOR_FRONT== WALL_ABSENT) next_DIRECTION = UP;
            else  next_DIRECTION = RIGHT;
        end
        UP: begin
            if (SENSOR_LEFT == WALL_ABSENT) next_DIRECTION = LEFT;
            else if (SENSOR_FRONT==WALL_ABSENT) next_DIRECTION = UP;
            else if (SENSOR_RIGHT==WALL_ABSENT) next_DIRECTION = RIGHT;
            else  next_DIRECTION = DOWN;
        end
        RIGHT: begin
            if (SENSOR_FRONT==WALL_ABSENT) next_DIRECTION = UP;
            else if (SENSOR_RIGHT==WALL_ABSENT) next_DIRECTION = RIGHT;
            else if (SENSOR_BACK ==WALL_ABSENT) next_DIRECTION = DOWN;
            else  next_DIRECTION = LEFT;
        end
        DOWN: begin
            if (SENSOR_RIGHT==WALL_ABSENT) next_DIRECTION = RIGHT;
            else if (SENSOR_BACK ==WALL_ABSENT) next_DIRECTION = DOWN;
            else if (SENSOR_LEFT ==WALL_ABSENT) next_DIRECTION = LEFT;
            else  next_DIRECTION = UP;
        end
        endcase
        MOTION_DIRECTION <= next_DIRECTION;
    end        
    endmodule
endmechanicalmodule
\end{lstlisting}

\section{MDL code for Example II}

\begin{lstlisting}
mechanicalmodule lock();
  //Geometry definition
  boundary {     //clockwise on the grid integer numbers
      line frontline1 {(0,35), (20, 49)},
      line frontline2 {(20, 49), (39, 35)},
      line rightline {(39, 35), (39,0)},
      line downline {(39,0), (0,0)},
      line leftline {(0,0), (0, 35)}};

  //Sensor definitions (name + location + value) 
  sensor pass_in {
      location = {(leftline , 0.2), (leftline , 0.4), (leftline , 0.6), (leftline , 0.8)},
      values = { "0b0": "OFF" , "0b1":  "ON"}};
  sensor action_btn {
      location = {(downline , 0.9)},
      values = { "0b0": "OFF" , "0b1":  "ON"}};
  sensor key {
      location = {(downline , 0.7)},
      values = { "0b0": "OFF" , "0b1":  "ON"}};

  actuator door {// Actuator
      type = "linear actuator",
      location = {(rightline , 0.5)},
      values = { "0b0": "Open", "0b1": "Closed"}};

    module lock_processor (
        input  wire       clk,
        input  wire [3:0] pass_in,     // 4-bit pass
        input  wire       action_btn,
        input  wire       key,         // 1=lock, 0=unlock
        output reg        door = 1'b0);  // 0=open, 1=locked
        
        localparam CLOSED = 1'b1, OPEN = 1'b0;
        reg [3:0] pass_memory = 4'b000;

        always @(posedge clk) begin
            // LOCK
            if (action_btn && key && door==OPEN) begin
                pass_memory[0] <= pass_in[0];  
                pass_memory[1] <= pass_in[1];
                pass_memory[2] <= pass_in[2];
                pass_memory[3] <= pass_in[3];// store pass
                door <= CLOSED;
            // UNLOCK
            end else if (action_btn && !key && door==CLOSED) begin
                if ((pass_memory[0] == pass_in[0]) &&
                    (pass_memory[1] == pass_in[1]) &&
                    (pass_memory[2] == pass_in[2]) &&
                    (pass_memory[3] == pass_in[3]))
                    door <= OPEN;
                else
                    door <= CLOSED;
            end
        end
    endmodule
endmechanicalmodule

\end{lstlisting}

\section{Design Automation with LLM (system prompt and generated MDL codes)}

\begin{figure}[htbp]
    \centering
    \includegraphics[width=\columnwidth]{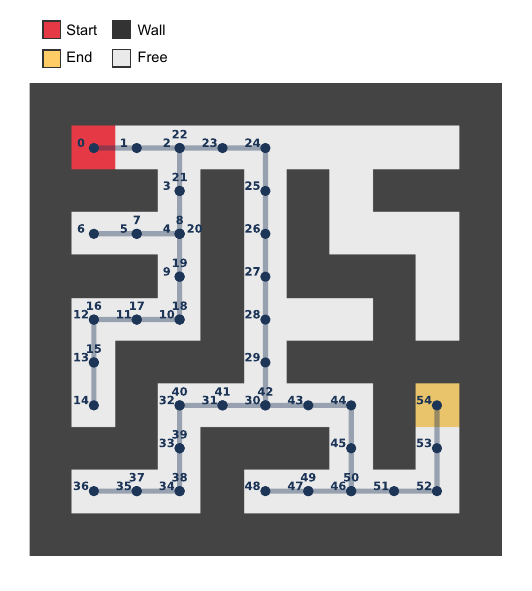}
    \caption{\textbf{Trajectory of the numerically simulated robot synthesized from LLM-generated MDL code using the right-wall-following navigation rule.}   } 
    \label{fig:API-Maze}
\end{figure}

For generating the MDL code with LLM, we first introduced the mechanical language and its components in the system prompt.
The prompt was refined a few times until it fully captured the specific characteristics of the language. The final version used to generate the MDL codes for the examples is presented below.

\lstdefinestyle{diff}{ basicstyle=\ttfamily\footnotesize, breaklines=true, 
frame=single, backgroundcolor=\color{gray!5},
breakindent=0pt, 
columns=flexible,
keepspaces=true, 
moredelim=**[is][\color{red}]{@@}{@@}
}

\UseRawInputEncoding
\begin{lstlisting}[style=diff]
MDL is a Verilog-like Mechanical Description Language, with which we define mechanical processors.
 
Key rules:

A mechanical description is wrapped in mechanicalmodule <name>(); ... endmechanicalmodule.

Boundary is defined as a polygon of line entries listed clockwise. 
Each line is named and uses two integer points: 
line name {(x1,y1),(x2,y2)}

Sensors and actuators are declared by name. Each has:
location = { (linename, position), ... } 

position is a proportion along the line measured from the line's first point; 
We use values from 0.0 to 1.0. 
Multiple (line, position) pairs are allowed for multiple physical placements. 
For any number of digits used in the input/output, the same number of locations should be indicated. 
Therefore, locations of different digits of the same sensor/actuators should not have the same values.

values = { "<binary>": "LABEL", ... } - maps encoded bits to human labels.

Actuators may include a type field, e.g. "directed actuator".


Inside the module you write Verilog-like code: 
module <name>(inputs..., outputs...); ... endmodule. 
We use always @(posedge clk) for state updates and case/if for behavioral logic. 
All the logic goes to this always @(posedge clk) block if the circuit is sequential.
Registeration happens at the end of the block not in all the if, or case blocks.

Sensors = inputs 
Actuators = outputs
Binary encoding width must match the value entries (e.g., "0b00" -> 2-bit actuator).

Always use clk as clock name.

Example:
mechanicalmodule <module_name>();

 boundary {
    line <name> {(x1,y1), (x2,y2)},
    ... };

 sensor <SENSOR_NAME> {
    location = {(lineName, pos), ...},
    values = { "0b0": "OFF", "0b1": "ON" } };

 actuator <ACTUATOR_NAME> {
    type = "<type_string>",
    location = {(lineName, pos), ...},
    values = { "0b00": "LABEL1", "0b01": "LABEL2", ... } };

 module <verilog_module_name>(
     input wire clk,
     input wire SENSOR_A,
     ...
     output reg [N:0] ACTUATOR_B );
     
    // Verilog-like behavioral code here
 endmodule

endmechanicalmodule

Don't initialize anything, we assume 0 for all initial conditions.
Based on these rules, I will ask you to write a MDL code for me that has a desired behavioral functionality. 
I want to use your respond directly so only respond with the MDL code with no other comments and don't write mdl on top.
\end{lstlisting}

Using the prompt given in section IV and the system prompt, the API model was able to produce a working MDL code for the case study presented in Example I, on the first attempt. The resulting algorithm implements the right-wall-following rule instead of the left variant for maze solving. The generated MDL code is shown below, and the corresponding simulated robot trajectory is presented in Fig.~\ref{fig:API-Maze}.

\begin{lstlisting}
mechanicalmodule maze_solver_robot();

 boundary {
    line left {(0,0), (0,39)},
    line top {(0,39), (39,39)},
    line right {(39,39), (39,0)},
    line bottom {(39,0), (0,0)} };

 sensor WALL_LEFT {
    location = {(left, 0.5)},
    values = { "0b0": "CLEAR", "0b1": "WALL" } };

 sensor WALL_UP {
    location = {(top, 0.5)},
    values = { "0b0": "CLEAR", "0b1": "WALL" } };

 sensor WALL_RIGHT {
    location = {(right, 0.5)},
    values = { "0b0": "CLEAR", "0b1": "WALL" } };

 sensor WALL_DOWN {
    location = {(bottom, 0.5)},
    values = { "0b0": "CLEAR", "0b1": "WALL" } };

 actuator MOVE {
    type = "directed actuator",
    location = {(bottom, 0.25), (bottom, 0.75)},
    values = { "0b00": "LEFT", "0b01": "UP", "0b10": "RIGHT", "0b11": "DOWN" } };

 module MazeSolver(
     input wire clk,
     input wire WALL_LEFT,
     input wire WALL_UP,
     input wire WALL_RIGHT,
     input wire WALL_DOWN,
     output reg [1:0] MOVE );

    localparam [1:0] DIR_LEFT  = 2'b00;
    localparam [1:0] DIR_UP    = 2'b01;
    localparam [1:0] DIR_RIGHT = 2'b10;
    localparam [1:0] DIR_DOWN  = 2'b11;

    reg [1:0] heading;
    reg [1:0] next_heading;
    reg [1:0] next_move;

    reg front, right_side, left_side, back_side;

    always @(posedge clk) begin
        next_heading = heading;

        case (heading)
            DIR_LEFT: begin
                front = WALL_LEFT;
                right_side = WALL_UP;
                left_side = WALL_DOWN;
                back_side = WALL_RIGHT;
            end
            DIR_UP: begin
                front = WALL_UP;
                right_side = WALL_RIGHT;
                left_side = WALL_LEFT;
                back_side = WALL_DOWN;
            end
            DIR_RIGHT: begin
                front = WALL_RIGHT;
                right_side = WALL_DOWN;
                left_side = WALL_UP;
                back_side = WALL_LEFT;
            end
            default: begin // DIR_DOWN
                front = WALL_DOWN;
                right_side = WALL_LEFT;
                left_side = WALL_RIGHT;
                back_side = WALL_UP;
            end
        endcase

        if (!right_side) begin
            case (heading)
                DIR_LEFT:  next_heading = DIR_UP;
                DIR_UP:    next_heading = DIR_RIGHT;
                DIR_RIGHT: next_heading = DIR_DOWN;
                default:   next_heading = DIR_LEFT;  // DIR_DOWN
            endcase
        end else if (!front) begin
            next_heading = heading;
        end else if (!left_side) begin
            case (heading)
                DIR_LEFT:  next_heading = DIR_DOWN;
                DIR_UP:    next_heading = DIR_LEFT;
                DIR_RIGHT: next_heading = DIR_UP;
                default:   next_heading = DIR_RIGHT; // DIR_DOWN
            endcase
        end else begin
            case (heading)
                DIR_LEFT:  next_heading = DIR_RIGHT;
                DIR_UP:    next_heading = DIR_DOWN;
                DIR_RIGHT: next_heading = DIR_LEFT;
                default:   next_heading = DIR_UP;    // DIR_DOWN
            endcase
        end

        next_move = next_heading;

        heading <= next_heading;
        MOVE <= next_move;
    end
 endmodule

endmechanicalmodule
\end{lstlisting}

With the exact prompts described here, the API model generated the MDL code for the case study presented Example II, on the first attempt. The resulting code correctly implemented the unlocking condition, yielding valid state transitions. The final MDL code is presented below:

\begin{lstlisting}
mechanicalmodule house_lock();

 boundary {
    line left_wall {(0,0), (0,35)},
    line left_roof {(0,35), (20,49)},
    line right_roof {(20,49), (39,35)},
    line right_wall {(39,35), (39,0)},
    line bottom {(39,0), (0,0)} };

 sensor KEYPAD {
    location = {(left_wall, 0.15), (left_wall, 0.35), (left_wall, 0.55), (left_wall, 0.75)},
    values = {
        "0b0000": "K0",  "0b0001": "K1",  "0b0010": "K2",  "0b0011": "K3",
        "0b0100": "K4",  "0b0101": "K5",  "0b0110": "K6",  "0b0111": "K7",
        "0b1000": "K8",  "0b1001": "K9",  "0b1010": "K10", "0b1011": "K11",
        "0b1100": "K12", "0b1101": "K13", "0b1110": "K14", "0b1111": "K15" } };

 sensor ACTION {
    location = {(bottom, 0.20)},
    values = { "0b0": "IDLE", "0b1": "PRESS" } };

 sensor MODE {
    location = {(bottom, 0.80)},
    values = { "0b0": "UNLOCK", "0b1": "LOCK" } };

 actuator DOOR {
    type = "directed actuator",
    location = {(right_wall, 0.50)},
    values = { "0b0": "OPEN", "0b1": "LOCKED" } };

 module lock_mechanism(
     input wire clk,
     input wire [3:0] KEYPAD,
     input wire ACTION,
     input wire MODE,
     output reg DOOR );
     
    reg [3:0] pwd;
    reg [3:0] next_pwd;
    reg next_door;

    always @(posedge clk) begin
        // Hold current state by default
        next_pwd  = pwd;
        next_door = DOOR;

        if (ACTION) begin
            if (MODE) begin
                // Locking: only when currently unlocked
                if (DOOR == 1'b0) begin
                    next_pwd  = KEYPAD;  // store password
                    next_door = 1'b1;    // lock
                end
            end else begin
                // Unlocking: only when currently locked
                if (DOOR == 1'b1) begin
                    if (KEYPAD == pwd)
                        next_door = 1'b0; // open
                    else
                        next_door = 1'b1; // stay locked
                end
            end
        end

        // Register updates at end
        pwd  <= next_pwd;
        DOOR <= next_door;
    end
 endmodule

endmechanicalmodule
\end{lstlisting}




\end{document}